\documentclass[aps,prl,twocolumn,floatfix]{revtex4} 
\usepackage[dvips]{graphicx} 
\usepackage{epsfig} 
\usepackage{pst-plot} 
\usepackage{bm} 
 
\begin{document} 
\title{Coupled cluster calculations of ground and excited states 
of nuclei} 
\author{K.~Kowalski$^1$, D.J. Dean$^2$, M.~Hjorth-Jensen$^3$, 
T.~Papenbrock$^{2,4}$, and P. Piecuch$^{1,5}$} 
\affiliation{$^1$Department of Chemistry, Michigan State University, 
East Lansing, MI 48824} 
\affiliation{$^2$Physics Division, Oak Ridge National Laboratory, 
P.O. Box 2008, Oak Ridge, TN 37831} 
\affiliation{$^3$Department of Physics and Center of 
Mathematics for Applications, University of Oslo, N-0316 Oslo, Norway} 
\affiliation{$^4$Department of Physics and Astronomy, University of Tennessee, 
Knoxville, TN 37996 USA} 
\affiliation{$^5$Department of Physics and Astronomy, 
Michigan State University, East Lansing, MI 48824} 
\date{\today} 
\begin{abstract} 
The standard and renormalized coupled cluster methods with singles, 
doubles, and noniterative triples and their generalizations to 
excited states, based on the equation of motion coupled cluster 
approach, are applied to the $^4$He and $^{16}$O nuclei. A comparison 
of coupled cluster results with the results of the exact 
diagonalization of the Hamiltonian in the same model space shows that 
the quantum chemistry inspired coupled cluster approximations provide 
an excellent description of ground and excited states of nuclei. The bulk 
of the correlation effects is obtained at the coupled cluster singles 
and doubles level. Triples, treated noniteratively, provide the 
virtually exact description. 
\end{abstract} 
 
\maketitle 
 
The description of finite nuclei requires an understanding of both 
ground- and excited-state properties based on a given nuclear 
Hamiltonian. While much progress has been made in employing the 
Green's Function Monte Carlo \cite{pieper02} and no-core shell model 
\cite{bruce2} techniques, 
these 
methods have apparent limitations to light nuclei. Given that present 
nuclear structure research facilities and the proposed Rare Isotope 
Accelerator will 
continue to 
open significant territory into regions 
of medium-mass and heavier nuclei, it becomes imperative to 
investigate methods that will allow for a description of medium-mass 
systems. Coupled cluster theory is a particularly promising candidate 
for such an endeavor due to its enormous success in quantum chemistry 
\cite{cizek66,cizek69,%
Bartlett95,Paldus99,comp_chem_rev00,Piecuch02a,Piecuch02b}. 
 
Coupled cluster theory originated in nuclear physics 
\cite{coester58,coester60} around 1960.  Early studies in the 
seventies \cite{kum78} probed ground-state properties in limited 
spaces with free nucleon-nucleon interactions available at the 
time. The subject was revisited
only recently by Guardiola {\it et al.}
\cite{bishop96}, for further theoretical development, and by Mihaila and
Heisenberg \cite{hm99}, for coupled cluster calculations
using realistic
two- and three-nucleon 
bare interactions
and expansions in the 
inverse particle-hole energy spacings.
However, much of 
the impressive development in 
coupled cluster theory made in quantum chemistry in
the last 15-20 years 
\cite{Bartlett95,Paldus99,comp_chem_rev00,Piecuch02a,Piecuch02b}
still awaits applications to the nuclear many-body problem.

In this Letter, we apply quantum chemistry inspired coupled cluster
methods
\cite{cizek66,cizek69,Bartlett95,Paldus99,comp_chem_rev00,%
Piecuch02a,Piecuch02b,Stanton:1993,Piecuch99} to
finite nuclei. We show that the coupled cluster approach is 
numerically inexpensive and accurate by comparing our results for 
$^{4}$He with results from exact diagonalization in a model space 
consisting of four major oscillator shells. For the first time, we 
apply coupled cluster theory to excited states 
in nuclei, exploiting the equation of motion coupled cluster formalism 
\cite{Stanton:1993,Piecuch99}. We 
discuss several approximations 
within coupled cluster theory and also compute the ground state of the 
$^{16}$O nucleus within the same model space. We remind the reader 
that certain acronyms have become standard in quantum chemistry. For 
this reason, we use the same abbreviations in this Letter. 

Coupled cluster theory is based on an exponential ansatz for the ground-state 
wave function 
$|\Psi_{0}\rangle=\exp(T) |\Phi\rangle$. 
Here $T$ is the cluster operator and $|\Phi\rangle$ is the reference 
determinant. In the CCSD (``coupled cluster with singles and 
doubles'') method, we truncate the many-body expansion of the cluster 
operator $T$ at two-body components. The truncated cluster operator 
$T^{\rm (CCSD)}$, used in the CCSD calculations, 
has the form \cite{purvis82}: $T^{\rm (CCSD)} = T_{1} 
+ T_{2}$.  Here $T_1=\sum_{i,a} t_a^i a^{a} a_{i}$ and 
$T_2= \frac{1}{4} \sum_{ij,ab} t_{ab}^{ij} a^{a} a^{b} a_{j} a_{i}$ are the singly 
and doubly excited clusters, with indices $i,j,k$ ($a,b,c$) 
designating the single-particle states occupied (unoccupied) in the 
reference Slater determinant $|\Phi\rangle$ and $a^{p}$ ($a_{p}$) 
representing the creation (annihilation) operators.  We determine the 
singly and doubly excited cluster amplitudes $t_a^i$ and 
$t_{ab}^{ij}$, defining $T_1$ and $T_2$, respectively, by solving the 
nonlinear system of algebraic equations, 
$\langle \Phi_{i}^{a} | \bar{H}^{\rm (CCSD)}|\Phi\rangle = 0$, 
$\langle \Phi_{ij}^{ab} | \bar{H}^{\rm (CCSD)}|\Phi\rangle = 0$, 
where 
$\bar{H}^{{\rm (CCSD)}} = \exp(-T^{\rm (CCSD)}) \, H \, \exp(T^{\rm (CCSD)})$ 
is the similarity-transformed Hamiltonian 
and $|\Phi_{i}^{a}\rangle$ and $|\Phi_{ij}^{ab}\rangle$ are 
the singly and doubly 
excited Slater determinants, respectively. 
Once $T_1$ and $T_2$ amplitudes are determined, we calculate the ground-state 
CCSD energy $E_{0}^{\rm (CCSD)}$ as 
$\langle\Phi|\bar{H}^{{\rm (CCSD)}}|\Phi\rangle $. 
For the excited 
states $|\Psi_{K}\rangle$ and energies $E_{K}^{\rm (CCSD)}$ ($K > 0$),
we apply the EOMCCSD (``equation of motion CCSD'') approximation, in which 
\begin{equation} 
|\Psi_{K}\rangle=R_{K}^{\rm (CCSD)} \exp(T^{\rm (CCSD)}) |\Phi\rangle . 
\label{eomfun} 
\end{equation} 
Here $R_{K}^{\rm (CCSD)} = R_{0}+ R_{1} + R_{2}$ is a sum of the 
reference ($R_{0}$), one-body ($R_{1}$), and two-body ($R_{2}$) 
components that are obtained by diagonalizing the 
similarity-transformed Hamiltonian $\bar{H}^{{\rm (CCSD)}}$ 
in the same space of singly and doubly excited determinants 
$|\Phi_{i}^{a}\rangle$ and $|\Phi_{ij}^{ab}\rangle$ as used in the 
ground-state CCSD calculations \cite{Stanton:1993,Piecuch99}. 
 
The CCSD and EOMCCSD methods are expected to describe the bulk of the 
correlation effects with inexpensive computational steps that scale as 
$n_{o}^{2} n_{u}^{4}$, where $n_{o}$ ($n_{u}$) is the number of 
occupied (unoccupied) single-particle orbitals. While the inclusion of 
triply excited clusters $T_{3}$ and three-body excitation operators 
$R_{3}$ increases the accuracy of the method, the 
resulting full CCSDT (``T'' stands for ``triples'') \cite{Noga:1987a} 
and EOMCCSDT \cite{Kowalski:2001d} methods scale 
as $n_{o}^{3} n_{u}^{5}$ and are rather 
expensive. For this reason, 
we add the {\it a posteriori} corrections due to triples 
to the CCSD/EOMCCSD energies, 
which require $n_{o}^{3} n_{u}^{4}$ noniterative steps. 
The 
ground- and excited-state triples corrections, $\delta_{0}$ and 
$\delta_{K}$ ($K>0$), respectively, are calcultated with the CR-CCSD(T) 
(``completely renormalized CCSD(T)'') approach 
\cite{Piecuch02a,Piecuch02b,Kowalski00,Kowalski03} in which 
\begin{equation} 
\delta_{K} = \mbox{$\frac{1}{36}$} \sum_{ijk,abc} \langle \tilde{\Psi}_{K} | \Phi_{ijk}^{abc} \rangle 
\, {\cal M}_{abc}^{ijk}(K) / \Delta_{K} \;\; (K \geq 0). 
\label{deltak} 
\end{equation} 
Here $|\Phi_{ijk}^{abc}\rangle$ are the triply excited 
determinants 
and ${\cal M}_{abc}^{ijk}(K)$ are 
the generalized moments of the CCSD ($K=0$) 
and EOMCCSD ($K > 0$) equations \cite{Kowalski00,Kowalski03,Kowalski01}, 
\begin{equation} 
{\cal M}_{abc}^{ijk}(K) = 
\langle \Phi_{ijk}^{abc} | \bar{H}^{{\rm (CCSD)}} 
S_{K}^{\rm (CCSD)} | \Phi \rangle \;, 
\label{mk} 
\end{equation} 
where $S^{\rm (CCSD)}_0=1$ and $S^{\rm (CCSD)}_K=R_K^{\rm (CCSD)}$ for $K > 0$. 
They can be calculated using the CCSD and EOMCCSD 
cluster and excitation operators $T^{\rm (CCSD)}$ and 
$R_{K}^{\rm (CCSD)}$, respectively. The $\Delta_{K}$ denominators are 
defined as 
\begin{equation} 
\Delta_{K}= 
\langle \tilde{\Psi}_{K} | S_{K}^{\rm (CCSD)} \exp(T^{\rm (CCSD)}) |\Phi\rangle \, . 
\label{denomk} 
\end{equation} 
The states $|\tilde{\Psi}_{K}\rangle$ in Eqs.~(\ref{deltak}) 
and (\ref{denomk}) include the leading triples contributions resulting 
from the perturbative 
analysis of the CCSDT and EOMCCSDT equations. We have 
$\mid\tilde{\Psi}_{0}\rangle=\bar{P}\exp(T^{\rm (CCSD)} + \tilde{T}_{3})|\Phi\rangle$ 
and $|\tilde{\Psi}_{K}\rangle = \bar{P}(R_{K}^{\rm (CCSD)} + \tilde{R}_{3}) 
\exp(T^{\rm (CCSD)}) |\Phi\rangle$ for $K > 0$, where $\bar{P}$ is a projection 
operator on the subspace spanned by the reference $|\Phi\rangle$ and singly, 
doubly, and triply excited determinants. 
The most complete forms of $\tilde{T}_{3}$ and $\tilde{R}_{3}$
defining the
\mbox{CR-CCSD(T),c}  approximation are 
\cite{Piecuch02a,Kowalski03} 
\begin{eqnarray} 
\tilde{T}_{3} &=& \mbox{$\frac{1}{36}$} 
\sum_{ijk,abc} ({\cal M}_{abc}^{ijk}(0)/D_{ijk}^{abc}(0) ) 
a^{a} a^{b} a^{c} a_{k} a_{j} a_{i}, 
\label{t3tilde} \\ 
\tilde{R}_{3} &=& \mbox{$\frac{1}{36}$} 
\sum_{ijk,abc} ({\cal M}_{abc}^{ijk}(K)/D_{ijk}^{abc}(K) ) 
a^{a} a^{b} a^{c} a_{k} a_{j} a_{i}, 
\label{r3tilde} 
\end{eqnarray} 
where $D_{ijk}^{abc}(K) = E_{K}^{\rm (CCSD)} - 
\langle \Phi_{ijk}^{abc} | \bar{H}^{\rm (CCSD)} |\Phi_{ijk}^{abc} 
\rangle$. 
In the case of the ground-state calculations, 
we also consider simplified variants of the CR-CCSD(T) 
theory, termed CR-CCSD(T),a and CR-CCSD(T),b. In the 
case of CR-CCSD(T),b, the perturbative denominator $D_{ijk}^{abc}(0)$ 
is replaced by $- \langle \Phi_{ijk}^{abc} | \bar{H}_{1}^{\rm (CCSD)} 
|\Phi_{ijk}^{abc} \rangle$, where $\bar{H}_{1}^{\rm (CCSD)}$ is the 
one-body part of $\bar{H}^{\rm (CCSD)}$. For CR-CCSD(T),a we replace 
$D_{ijk}^{abc}(0)$ by the standard many-body perturbation theory 
(MBPT) triples denominator ($\epsilon_{i} + \epsilon_{j} + 
\epsilon_{k} - \epsilon_{a} - \epsilon_{b} - \epsilon_{c}$), where 
$\epsilon_{i}$ and $\epsilon_{a}$ are the diagonal elements of the 
Fock matrix.  Very accurate results for the excitation energies $E_{K} 
- E_{0}$ of many-electron systems are obtained if we use the complete 
CR-CCSD(T),c theory to calculate the energies of excited states and 
the CR-CCSD(T),b approximation for the ground-state energy 
\cite{Kowalski03}. For the ground states, it may sometimes be 
worthwhile to replace the $\Delta_{0}$ denominator, 
Eq. (\ref{denomk}), which renormalizes the triples correction 
$\delta_{0}$ by 1, since $\Delta_{0}$ equals 1 plus terms of the 
second MBPT order or higher \cite{Kowalski00}.  We indicate this by 
using acronyms, such as CR-CCSD(T),c/$\Delta_{0}=1$ (as opposed to 
CR-CCSD(T),c, where $\Delta_{0}$ is included). 
 
We use the Idaho-A nucleon-nucleon potential 
\cite{machleidt02} which was produced using
techniques of chiral 
effective field theory \cite{weinberg,vankolck}. 
Modern two-nucleon interactions, such as Idaho-A, include short-range 
repulsive cores that require calculations in extremely large model 
spaces to reach converged results \cite{hm99}. In order to remove the 
hard-core part of the interaction from the problem and thereby allow 
for realistic calculations in manageable model spaces, we renormalize 
the interactions through a $G$-matrix procedure for use in the 
$0s$-$0p$-$0d1s$-$0f1p$ oscillator basis. Our Hamiltonian is thus 
given by $H=t+G(\tilde{\omega})$, where $\tilde\omega$ is the 
$G$-matrix starting energy. We use a simple procedure described in 
Ref.~\cite{herbert02} to alleviate the starting-energy dependence of 
the $G$-matrix in orbitals below the Fermi surface. We also modify 
the Hamiltonian by adding to it the center-of-mass Hamiltonian times a 
Lagrange multiplier $\beta_{\rm c.m.}$.  Thus, our Hamiltonian becomes 
$H^{\prime}=H+\beta_{\rm c.m.}H_{\rm c.m.}$. We choose $\beta_{\rm 
c.m.}$ such that the expectation value of $H_{\rm c.m.}=0.0$~MeV. 
Details may be found in Ref.~\cite{dean03}. 
 
We tested the performance of the above coupled cluster approximations 
in the context of the nuclear many-body problem by applying them to two 
closed-shell nuclei, $^{4}$He and $^{16}$O, in the one-particle space of 
four major oscillator shells. Shell model diagonalization provided an 
exact answer for a given Hamiltonian in the $^4$He case.  Comparing 
the exact ground- and excited-state energies resulting from the 
diagonalization of the Hamiltonian in the small model space with the 
coupled cluster energies obtained in the same model space, we can assess 
the usefulness of various coupled cluster approximations in 
calculations for atomic nuclei. In particular, we can learn about the 
possible role of triply excited clusters in an accurate description of 
ground and excited states {\it without confusing the inaccuracies 
resulting from the inadequate treatment of the many-body problem by a 
given coupled cluster approximation with other sources of error}. 
 
We report our results for the ground-state energy of $^4$He in 
Table~\ref{table_he4_gs}. We used two types of reference determinants 
$|\Phi\rangle$: one constructed from the lowest-energy oscillator 
states and the Hartree-Fock determinant. Throughout the table we see 
that the results obtained in the oscillator basis are lower in energy 
when compared to those obtained in the Hartree-Fock basis. The two 
best methods in the oscillator basis are
CR-CCSD(T),a/$\Delta_{0}=1$ and CR-CCSD(T),c/$\Delta_{0}=1$.
They yield results within $40$~keV and $300$~keV of the full configuration interaction
(CI) diagonalization problem, respectively. The 
CR-CCSD(T),a/$\Delta_{0}=1$ approach applied to the oscillator basis 
overshoots the exact result, which is a consequence of using the 
standard MBPT denominators in the definition of $\tilde{T}_{3}$, 
Eq. (\ref{t3tilde}).  The CR-CCSD(T),a approach, in which the triples 
correction $\delta_{0}$ is renormalized via the presence of the 
$\Delta_{0}$ denominator in Eq. (\ref{deltak}), is more stable in this 
regard, providing the upper bound to the energy, although the 600 keV 
error obtained with the CR-CCSD(T),a method is not as impressive as 
the $40$~keV error obtained with CR-CCSD(T),a/$\Delta_{0}=1$.  In 
general, the CR-CCSD(T) results are considerably more accurate than 
the results of the CCSD calculations, in which $T_{3}$ is ignored, 
although the CCSD approach describes the bulk of the correlation effects, 
reducing the large 16.273 MeV error obtained by calculating $\langle 
\Phi | H | \Phi \rangle$ with the oscillator reference $| \Phi 
\rangle$ to 1.5 MeV.  The effectiveness of the CCSD approach can also 
be illustrated by comparing the CCSD energy with the results of 
truncated shell-model calculations (CISD), in which the Hamiltonian is 
diagonalized in the same space of singly and doubly excited 
determinants as used in the CCSD caculations. The costs of 
the CISD and CCSD calculations are almost identical (both are 
$n_{o}^{2} n_{u}^{4}$ procedures), and yet the error in the CISD energy 
obtained in the oscillator basis is twice as large as the error 
obtained with CCSD.  The noniterative triples corrections defining the 
CR-CCSD(T) approaches reduce these errors to as little as $40$~keV, 
which is a lot better than the 1.3 MeV error in the CISDT 
calculations, where the Hamiltonian is diagonalized in the much larger 
space of all singly, doubly, and triply excited determinants.  This 
demonstrates the advantages of coupled cluster methods over the 
diagonalization techniques. Similar observations apply to 
the Hartree-Fock basis, although the coupled cluster results obtained 
with this basis are not as good as those obtained with the 
oscillator basis. For example, the best result in the Hartree-Fock 
basis, obtained with CR-CCSD(T),c/$\Delta_{0}=1$, is $700$~keV above 
the exact result. This suggests that we may be better off by 
using the oscillator basis in coupled cluster calculations. On the 
other hand, the CR-CCSD(T) results obtained in the Hartree-Fock basis 
are not unreasonable, allowing us to contemplate the use of the 
Hartree-Fock basis in coupled cluster calculations for 
open-shell nuclei. (This would parallel the Hartree-Fock-based coupled 
cluster calculations for open-shell 
electronic states in 
chemistry.) 
 
\begin{table} 
\caption{The ground-state energies of $^4$He 
calculated using 
the oscillator (Osc) and Hartree-Fock (HF) basis states. 
Units are MeV. 
The reference energies $\langle \Phi | H^{\prime} | \Phi \rangle$ 
are -7.211 (Osc) and -10.520 (HF) MeV. 
} 
\begin{center} 
\begin{tabular}{|ccc|} 
\hline 
Method & Osc & HF \cr 
\hline 
CCSD                       & -21.978 & -21.385 \cr 
CR-CCSD(T),a           & -22.841 & -22.395 \cr 
CR-CCSD(T),a/$\Delta_{0}=1$ & -23.524 & -22.711 \cr 
CR-CCSD(T),b           & -22.396 & -22.179 \cr 
CR-CCSD(T),b/$\Delta_{0}=1$ & -22.730 & -22.428 \cr 
CR-CCSD(T),c           & -22.630 & -22.450 \cr 
CR-CCSD(T),c/$\Delta_{0}=1$ & -23.149 & -22.783 \cr 
CISD                       & -20.175 & -20.801 \cr 
CISDT                      & -22.235 &    --   \cr 
Exact                      & -23.484 & -23.484 \cr 
\hline 
\end{tabular} 
\end{center} 
\label{table_he4_gs} 
\end{table} 
 
We used the EOMCCSD method and its CR-CCSD(T) extension to compute 
excited states. To our knowledge, this is the first time that nuclear 
excited states are computed using coupled cluster methods. The 
results for $^4$He are given in Table~\ref{table_2}. The low-lying 
$J=1$ state most likely results from the center-of-mass contamination 
which we have removed only from the ground state.  The $J=0$ and $J=2$ 
states calculated using EOMCCSD and CR-CCSD(T) are in excellent 
agreement with the exact results.  For these two states, the EOMCCSD 
approach provides the relatively small, 0.3-0.4 MeV, errors, which are 
further reduced by the CR-CCSD(T) triples corrections to $< 0.1$ 
MeV. Based on the experience with the equation of 
motion coupled cluster methods in chemistry 
\cite{Stanton:1993,Bartlett95,Piecuch02a,Kowalski03,Kowalski01,Kowalski:2001d}, 
the very good performance of the EOMCCSD approach for the 
lowest-energy excited states of $^4$He can be understood if we realize 
that these states are dominated by single-particle excitations. 
Again, a comparison of the EOMCCSD and CISD results shows that coupled cluster 
theory offers much higher accuracies compared to truncated 
diagonalization of similar numerical effort. According to the 
experiment, the lowest lying $0^+$ state in $^4$He is a resonance at 
an excitation of 20.21~MeV and a width of 0.5~MeV, while the first 
$J=2$ state lies at 21.84~MeV and has a width of 2~MeV.  We have not 
identified the parity of our calculated states, but it seems to us 
that we will be able to model the excitations in $^4$He and other 
nuclei using coupled cluster theory.  Our results in 
Table~\ref{table_2} are indicative of the accuracies we may expect 
from such calculations. 
 
\begin{table} 
\caption{The excitation energies of $^4$He 
calculated using the 
oscillator basis states (in MeV). 
} 
\begin{center} 
\begin{tabular}{|ccccc|} 
\hline 
State & EOMCCSD & CR-CCSD(T)$^a$ & CISD & Exact \cr 
\hline 
J=1   &  11.791 & 12.044 & 17.515    & 11.465 \cr 
J=0   &  21.203 & 21.489 & 24.969    & 21.569 \cr 
J=2   &  22.435 & 22.650 & 24.966    & 22.697 \cr 
\hline 
\end{tabular} 
\end{center} 
$^a$ The difference of the CR-CCSD(T),c energy of excited state and 
the CR-CCSD(T),b energy of the ground state. 
\label{table_2} 
\end{table} 
 
We also applied the CCSD and CR-CCSD(T) methods to $^{16}$O.
Table~\ref{table_ox16_gs} shows the total ground-state energy values
obtained with the CCSD and CR-CCSD(T) approaches. As in the $^{4}$He
case, coupled cluster methods recover the bulk of the correlation
effects, producing the results of the CISDTQ, or better, quality.
CISDTQ stands for the expensive shell-model diagonalization in a huge space
spanned by the reference and all singly, doubly, triply, and quadruply
excited determinants (the most expensive steps of CISDTQ
scale as $n_{o}^{4} n_{u}^{6}$).
To understand this result, we note that
the CCSD $T_1$ and $T_2$ amplitudes are similar in order of magnitude. (For
an oscillator basis, both $T_1$ and $T_2$ contribute to the first-order
MBPT wave function.)
Thus, the $T_1 T_2$ {\it disconnected} triples are large, much larger than
the $T_3$ {\it connected} triples, and the difference
between the CISDT (CI singles, doubles, and triples)
and CISD energies is mostly due to $T_1 T_2$.
The small $T_3$ effects, as estimated by CR-CCSD(T), are consistent
with the CI diagonalization calculations. If the $T_3$ corrections 
were large, we would observe a significant lowering of the 
CCSD energy, far below the CISDTQ result.
The CISDTQ diagonalization is not size-extensive, while the
CCSD and CR-CCSD(T)/$\Delta_{0}=1$ methods maintain this property.
Moreover, the CCSD and CR-CCSD(T) methods
bring the nonnegligible higher-than-quadruple excitations, 
such as $T_1^3 T_2$, $T_1 T_2^2$, and $T_{2}^{3}$, which are 
not present in CISDTQ. It is, therefore, quite likely that the 
CR-CCSD(T) results are very close to the results of the exact
diagonalization, which cannot be performed.

\begin{table} 
\caption{The ground-state energy of $^{16}$O 
calculated using various coupled cluster methods 
and oscillator basis states. 
The reference energy $\langle \Phi | H^{\prime} | \Phi \rangle$ 
is -109.452 MeV. 
} 
\begin{center} 
\begin{tabular}{|cc|} 
\hline 
Method & Energy \cr 
\hline 
CCSD                       & -139.310 \cr 
CR-CCSD(T),a           & -139.465 \cr 
CR-CCSD(T),a/$\Delta_{0}=1$ & -139.621 \cr 
CR-CCSD(T),b           & -139.375 \cr 
CR-CCSD(T),b/$\Delta_{0}=1$ & -139.440 \cr 
CR-CCSD(T),c           & -139.391 \cr 
CR-CCSD(T),c/$\Delta_{0}=1$ & -139.467 \cr 
CISD                        & -131.887 \cr 
CISDT                       & -135.489 \cr 
CISDTQ                      & -138.387 \cr 
\hline 
\end{tabular} 
\end{center} 
\label{table_ox16_gs} 
\end{table} 
 
In summary, we used the quantum chemistry inspired coupled cluster 
approximations to calculate the ground and excited states of the 
$^{4}$He and $^{16}$O nuclei. By comparing coupled cluster results 
with the exact results obtained by diagonalizing the Hamiltonian in 
the same model space, we demonstrated that relatively inexpensive 
coupled cluster approximations recover the bulk of the nucleon 
correlation effects in ground- and excited-state nuclei. These results 
are a strong motivation to further develop coupled cluster methods for 
the nuclear many-body problem, so that accurate {\it ab initio} 
calculations for small- and medium-size nuclei become as routine as the 
molecular electronic structure calculations.

\begin{acknowledgments} 
Supported by the U.S. Department of Energy
under
Contract Nos. DE-FG02-96ER40963 (University of Tennessee),
DE-AC05-00OR22725 with UT-Battelle, LLC (Oak Ridge
National Laboratory), DE-FG02-01ER15228 (Michigan State University),
the Research Council of Norway, and the Alfred P. Sloan Foundation. 
\end{acknowledgments}

\end{document}